# Liquid Metal Oxide-assisted Integration of High-k Dielectrics and Metal Contacts for Two-Dimensional Electronics

*Dasari Venkatakrishnarao[1], Abhishek Mishra[1], Yaoju Tarn[1], Michel Bosman[1,2], Rainer Lee[1], Sarthak Das[1], Subhrajit Mukherjee[1], Teymour Talha-Dean[1,3], Yiyu Zhang[1], Siew Lang Teo[1], Jian Wei Chai[1], Fabio Bussolotti[1], Kuan Eng Johnson Goh\*[1,4,5], Chit Siong Lau\*[1,6]*

[1] Institute of Materials Research and Engineering, Agency for Science, Technology and Research (A*STAR), 2 Fusionopolis Way, Innovis, 138634, Singapore

[2] Department of Materials Science and Engineering, National University of Singapore, 9 Engineering Drive 1, 117575, Singapore

[3] Department of Physics and Astronomy, Queen Mary University of London, London, E14NS, United Kingdom

[4] Department of Physics, National University of Singapore, 2 Science Drive 3, 117551, Singapore

[5] Division of Physics and Applied Physics, School of Physical and Mathematical Sciences, Nanyang Technological University, 50 Nanyang Avenue 639798, Singapore

[6] Science, Mathematics and Technology, Singapore University of Technology and Design, 8 Somapah Road, 487372, Singapore

*Email: aaron_lau@imre.a-star.edu.sg; kejgoh@yahoo.com

**Abstract**

Two-dimensional van der Waals semiconductors are promising for future nanoelectronics. However, integrating high-k gate dielectrics for device applications is challenging as the inert van der Waals material surfaces hinder uniform dielectric growth. Here, we report a liquid metal oxide-assisted approach to integrate ultrathin, high-k $HfO_2$ dielectric on 2D semiconductors with atomically smooth interfaces. Using this approach, we fabricated 2D $WS_2$ top-gated transistors with subthreshold swings down to 74.5 mV/dec, gate leakage current density below $10^{-6}$ $A/cm^2$, and negligible hysteresis. We further demonstrate a one-step van der Waals integration of contacts and dielectrics on graphene. This can offer a scalable approach toward integrating entire prefabricated device stack arrays with 2D materials. Our work provides a scalable solution to address the crucial dielectric engineering challenge for 2D semiconductors, paving the way for high-performance 2D electronics.

The invention of the transistor over a half-century ago was a significant technological breakthrough that accelerated the explosive growth of information technology. Much of modern technological developments can be traced to the role of semiconductor transistors in electronics. Such rapid progress was enabled by Dennard's scaling laws[1] resulting in successful voltage and dimension downscaling of silicon-based complementary metal-oxide-semiconductor (CMOS) technology. The downscaling improved CMOS performance, cost, and energy consumption and paved the way for the growth of CMOS technology projected by the well-known Moore's law.[2] However, downscaling is approaching its fundamental limits; preserving gate electrostatics when reducing the overall physical device size will require a concurrent reduction in channel thickness.[3–5] However, device performance can degrade when the thickness of silicon is reduced due to increased carrier scattering from surface roughness and dangling bonds.[6]

Therefore, the 'More than Moore' concepts proposed by the International Roadmap for Devices and Systems (IRDS) look beyond silicon at emerging materials.[7–9] Two-dimensional (2D) semiconducting transition metal dichalcogenides (TMDs) are an exciting new class of materials for future electronics.[7,10–16] An attractive advantage of 2D TMDs is their layered van der Waals (vdW) structure, which enables the isolation of monolayers with sub-nm thickness and pristine surfaces lacking dangling bonds. Thus, unlike silicon, 2D TMD field effect transistors (FETs) can maintain high carrier mobility even at the monolayer limit. Another advantage of 2D TMDs is the wide range of materials with different mechanical, electronic, and optical properties. This is potentially useful for integrating devices with diverse functionalities, including flexible electronics, sensors, and optoelectronics tailored for specific applications, such as artificial intelligence and 'The Internet of Things.'[5,11,17]

However, dielectric integration remains a challenge in realizing 2D TMD electronics.[11,18,19] The gate dielectric plays a crucial role in the performance of energy-efficient low-power (LP) devices. Device figures of merit, such as gate leakage current and gate hysteresis, should be minimized for energy efficiency and reliability.[20,21] The subthreshold swing (*SS*) quantifies the steepness of the transistor turn-on characteristics and is the gate voltage required to increase the transistor drain current by an order of magnitude. *SS* should also be minimized; the theoretical thermionic limit is set at 60 mV/dec at room temperature.[22] However, in practice, devices are limited by the dielectric and interface quality. Attaining small *SS* requires a high gate capacitance, which can be achieved by reducing dielectric thickness, but thinner dielectrics can increase gate leakage currents. For example, exfoliated hexagonal boron nitride (hBN) is a commonly used dielectric in laboratories that can be integrated with atomically smooth interfaces. However, the low dielectric constant of hBN is insufficient for thickness scaling while preserving acceptable gate leakage currents.[23] Furthermore, micron-sized exfoliated hBN is intrinsically unscalable. Scalability is also challenging for other vdW dielectrics that have been explored, such as mica,[24] $CaF_2$,[25,26] $MoO_3$,[27] and $SrTiO_3$ perovskites.[28,29]

Alternatively, high-k dielectrics can be deposited, for example, in Intel's 14 nm nodes, where $HfO_2$ is integrated via atomic layer deposition (ALD).[30–32] However, ALD is difficult on 2D materials as their pristine surfaces do not promote growth precursor nucleation.[33] ALD-grown dielectrics on 2D materials are typically non-uniform, thin films contain pinholes, and interfaces are rough and defective.[34,35] The resulting devices have high gate leakage currents, SS values, and gate hysteresis. Surface modification, such as plasma, UV ozone, or metal seeding layers, can promote precursor nucleation.[32,36–38] However, these techniques damage or alter the 2D material surface and are especially severe in monolayer devices. Seed buffer layers of molecules and polymers can also effectively induce growth nucleation but typically have low dielectric constants that degrade the gate capacitance and can be thermally unstable.[39,40]

An emerging strategy is to deposit a dielectric on a sacrificial layer.[41,42] This sacrificial layer is then etched away to release the dielectric film, which can then be arbitrarily transferred onto a target substrate or over a 2D material of choice. Polymers or graphene can be used as sacrificial layers to integrate dielectrics with 2D TMDs. However, approaches demonstrated so far have restrictions in deposition technique choices and growth conditions due to the limitations of the sacrificial layer. Graphene as a sacrificial layer is incompatible with ALD and is limited to physical deposition techniques like evaporation. Polymers have low thermal stability, limiting growth temperatures and lowering material quality.

Here, we demonstrate the transfer of high-k ALD $HfO_2$ dielectric using a liquid metal (LM) oxide sacrificial layer. Large-area ultrathin films can be readily transferred without cracks and wrinkles, and atomically smooth interfaces with 2D TMDs can be achieved. Amongst TMDs, $MoS_2$ is the most studied material due to the availability of high-quality crystals. The less explored $WS_2$ can potentially offer higher carrier mobility due to its smaller effective mass (~0.3 $m_0$) and has attracted interest from industry such as IMEC and TSMC.[43–45] This work explores integrating $HfO_2$ with 2D $WS_2$ to form top-gated field effect transistors. The high-quality dielectric integration results in devices with low gate leakage currents < $10^{-6}$ $A/cm^2$ well below the IRDS LP limit, negligible gate hysteresis, and record low subthreshold swings of 74.5 mV/dec. We further demonstrate the versatility of our technique with a one-step transfer of both metal contacts and ALD $HfO_2$ dielectric to form graphene transistors, paving the way toward integrating entire device stack arrays.

Ultrathin LM gallium oxide is the ideal sacrificial layer for transferring dielectrics and metals in many ways. Unlike 2D vdW material surfaces, the hydrophilic gallium oxide surface with dangling bonds is excellent for promoting ALD precursor nucleation.[46] Furthermore, compared to polymer and graphene-based sacrificial layers, gallium oxide is robust against harsh processes and conditions such as oxygen plasma, UV ozone, and high temperatures above 500 °C (Figure S1).[47,48] This robustness is crucial for practical versatility and flexibility. A wider range of growth techniques, material choices, and substrate treatment processes can be used to precisely tailor the properties of the subsequently grown material for specific applications.[49] The gallium oxide-based sacrificial layer can also benefit the low-temperature back-end of line (BEOL) processes in developing an integrated circuit. It can potentially enable the integration of materials pre-processed at high temperatures, which otherwise cannot be integrated as a BEOL component due to limitations on thermal budget (400 °C). Moreover, this technique can potentially enable integration of various combinations of high-k dielectrics and 2D materials. Such integration can simplify efforts at improving device reliability by finding optimal combinations with respect to defect band alignments, which may otherwise not be possible using conventional direct ALD due to growth limitations.[50] Next, LM gallium oxides can be printed over large areas (> cm sizes) with excellent uniformity and are ultrathin (~3 nm) and atomically smooth (< 0.3 nm) due to the self-limiting nature of the Cabrera-Mott oxidation mechanism.[48,51,52] Atomically smooth growth surfaces ensure that ALD-grown materials are uniformly smooth and flat, as ALD is a conformal layer-by-layer growth process. The ultrathin nature of LM gallium oxide as a sacrificial layer reduces excess strain on the grown materials during the thermal cycling of ALD growth that can introduce unwanted cracks and wrinkles.[41] Finally, gallium oxide exhibits excellent etch selectivity with metals and $HfO_2$, which facilitates its easy delamination for transfer.

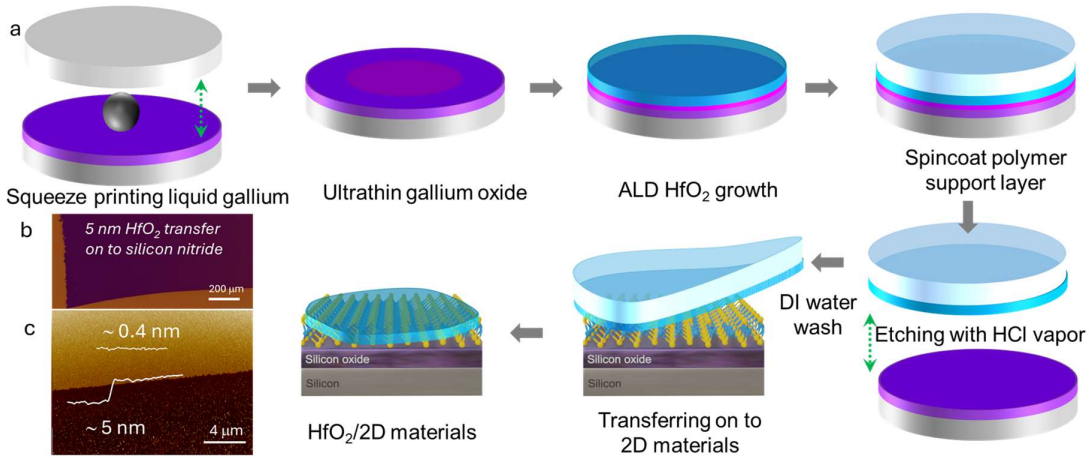

**Figure 1 Schematic illustration of liquid metal oxide-assisted $HfO_2$ transfer.** (a) First, ultrathin gallium oxide (~ 3 nm) is printed from liquid gallium metal onto a supporting substrate ($SiO_2$/Si). The gallium oxide surface is then subjected to $O_2$ plasma treatment to remove residue and improve its hydrophilicity before atomic layer deposition (ALD) of $HfO_2$. A polymer support layer is spin-coated for mechanical support before etching away the sacrificial gallium oxide layer with HCl vapor and rinsed with DI water. The $HfO_2$/polymer stack is then transferred onto a target substrate over 2D materials. (b) Optical microscope image of a large-

area transferred HfO$_2$ and its (c) atomic force microscopy profile showing a thickness of ~5 nm and a surface roughness down to 0.4 nm.

Figure 1a illustrates this LM-assisted transfer technique for ALD dielectrics. First, we print the sacrificial gallium oxide film from liquid gallium metal onto SiO$_2$, which acts as a supporting substrate, following the procedure described in our previous work.[48] Next, a short oxygen plasma treatment is performed to improve the hydrophilicity and cleanliness of the gallium oxide surface in preparation for ALD growth. We grow HfO$_2$ via ALD at a temperature of 200 °C; its thickness can be precisely controlled through the number of ALD cycles. We then prepare the film for transfer by spin coating a polymer support layer, followed by selective etching of the sacrificial gallium oxide layer using HCl vapor. The polymer/HfO$_2$ stack is then released and can be transferred onto another target substrate. Large area HfO$_2$ films up to cm sizes can be achieved (Figure 1b). In principle, even larger films approaching wafer-scale should be possible, which is promising for scalability; we are restricted mainly by equipment limitations, i.e., the size of the LM printing stage.

To first assess the material quality of our HfO$_2$ films, we perform atomic force microscopy (AFM) and X-ray photoemission spectroscopy (XPS). Our AFM analysis confirms that uniform and smooth HfO$_2$ films can be transferred with a root-mean-square (RMS) roughness of ~0.4 nm and a thickness down to 5 nm (Figure 1c). The highly stoichiometric HfO$_2$ films confirmed by XPS (Figure S6) affirm the suitability of LM gallium oxide as a growth substrate for ALD of thin film dielectrics. ALD growth of HfO$_2$ at 200 °C generally results in amorphous films.[49] For electronic applications, especially gate dielectrics, amorphous films are preferred over polycrystalline films due to less spatial variations in dielectric properties and suppressed leakage.[53]

Next, we investigate the interface quality between our transferred HfO$_2$ films and 2D materials. Figure 2a shows an optical image of a mechanically exfoliated 2D WS$_2$ flake with mono- and multi-layered regions. Atomic force microscopy analysis conducted on the flake performed before (Figure 2b) and after (Figure 2c) HfO$_2$ transfer showcases the conformal nature of our dielectric transfer technique. The solid white lines indicate the height profiles measured across the WS$_2$ edges where we observe no significant changes in step heights after HfO$_2$ transfer, suggesting that the WS$_2$/HfO$_2$ interface remains atomically smooth. This atomic smoothness of the interface is corroborated by cross-sectional scanning transmission electron microscopy imaging, which directly visualizes the interface morphology for both monolayer (Figure 2d) and multi-layered WS$_2$ (Figure 2e).

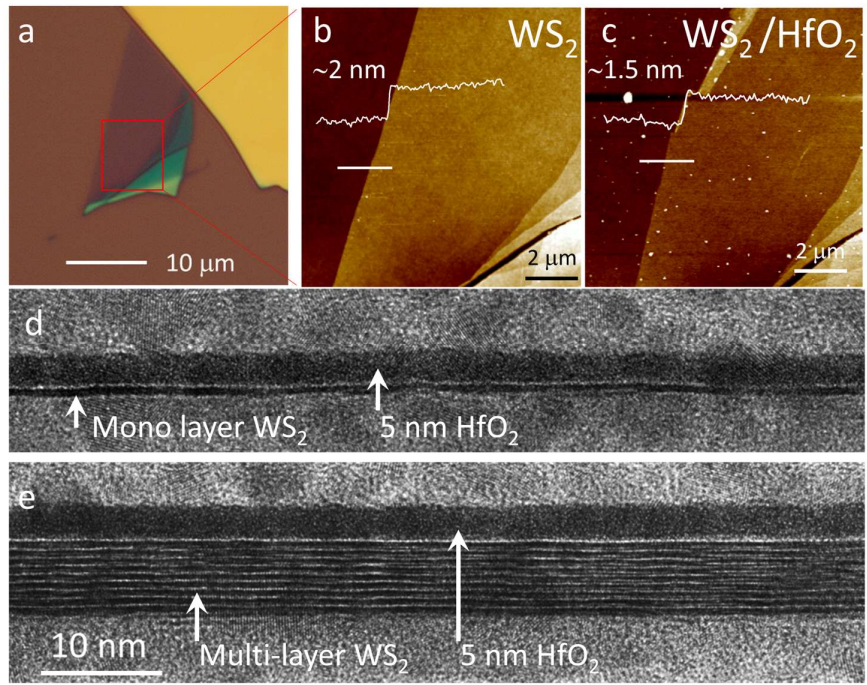

**Figure 2 Interface morphology**. (a) Optical image of an exfoliated WS$_2$ flake with monolayer and multilayer regions. Atomic force microscopy image of the WS$_2$ flake (b) before and (c) after HfO$_2$ transfer (thickness ~5 nm). The white solid lines indicate where the height profiles are taken. Cross-sectional scanning transmission electron microscopy image (d) mono- and (e) multi-layered WS$_2$/HfO$_2$ interface showing an atomically smooth interface.

Such excellent interfacial morphologies should translate to superior electrical characteristics. To assess the electrical properties of transferred HfO$_2$, we fabricated metal-insulator-metal (MIM) capacitors by transferring HfO$_2$ onto few-layer graphene/graphite electrodes (Figure 3a). By choosing graphite over conventional metals, we can achieve a sufficiently thin and smooth electrode to minimize mechanical strain in transferred HfO$_2$, which can be exacerbated when the dielectric film is draped over bulky electrode edges. Such unwanted strain can lead to higher leakage currents or even complete tears and would fail to provide an accurate assessment of our dielectric properties. For comparison, we fabricated MIM capacitors by directly growing ALD HfO$_2$ on Palladium (Pd) electrodes. With our MIM capacitors, we conducted current-voltage (*I-V*) and capacitance-voltage (*C-V*) measurements (Figure 3b, c). While we observe a slight increase in leakage currents for transferred films compared to as-grown films, overall, the leakage characteristics are excellent (<

$10^{-4}$ A/cm$^2$) and remain well below the IRDS requirement for low-power devices ($10^{-2}$ A/cm$^2$).[5] Breakdown characteristics are also comparable, at about 6 MV/cm (11 nm) and 7.6 MV/cm (16 nm) for transferred films versus 5.7 MV/cm (11 nm) and 8.7 (16 nm) for direct ALD-grown HfO$_2$ films. These results affirm the excellent insulating nature of our transferred HfO$_2$ films.

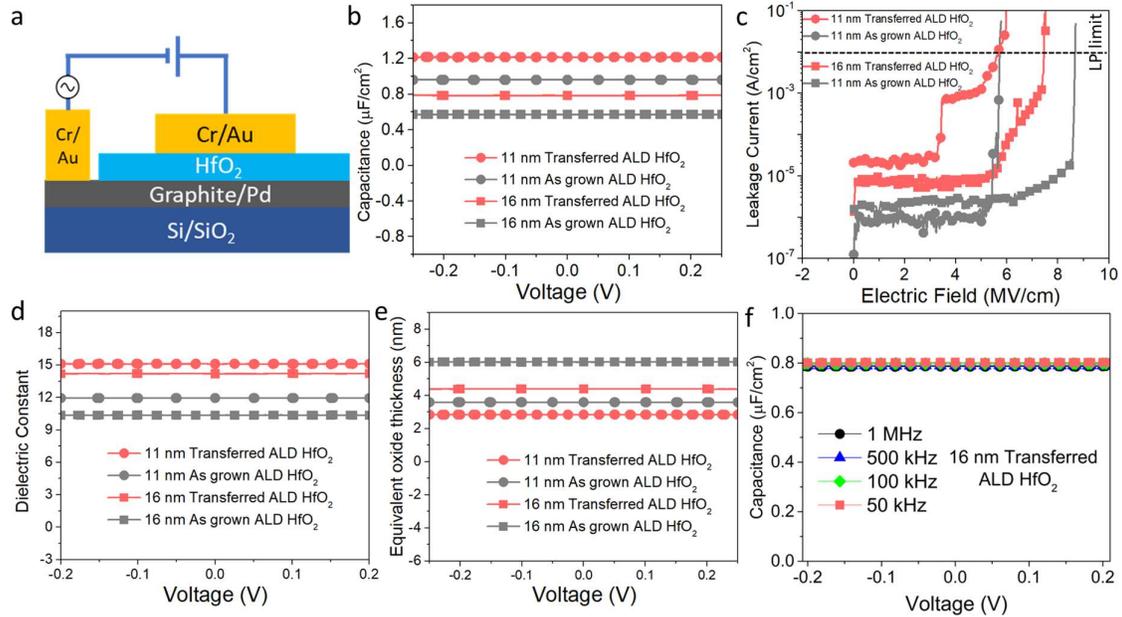

**Figure 3 Electrical properties of metal-insulator-metal (MIM) capacitors**. (a) Schematic of our MIM devices. For graphite devices, HfO$_2$ is transferred using the liquid metal-assisted technique. For Pd devices, HfO$_2$ was directly grown using ALD for comparison. (b) Capacitance-voltage measurements for 11 nm and 16 nm transferred and as-grown HfO$_2$ films. (c) Breakdown characteristics of 11 and 16 nm transferred and as-grown HfO$_2$ films. (d) Dielectric constants and (e) equivalent oxide thicknesses are extracted for 11 nm and 16 nm transferred and as-grown HfO$_2$ films. (f) Capacitance-voltage measurements for the 16 nm transferred HfO$_2$ device showed only slight variation at different measurement frequencies from 50 kHz to 1 MHz.

From the *C-V* measurements, we find that the capacitance values for our transferred HfO$_2$ films (11 nm, 1.2 µF/cm$^2$ and 16 nm, 0.8 µF/cm$^2$) are slightly higher than their direct ALD-grown counterparts (11 nm, 1.0 µF/cm$^2$ and 16 nm, 0.6 µF/cm$^2$), further evidence that the integrity and quality of our films are not compromised from the transfer process. High-quality interfaces with low trap densities are confirmed from our *C-V* measurements, where we find small frequency dispersion across a range of frequencies from 50 kHz to 1 MHz with less than 5% variation in capacitance values.[54]

We next calculate key dielectric parameters such as the dielectric constant $\varepsilon_{HfO2}$ = $C_{HfO2} t_{HfO2}/\varepsilon_0$ and the equivalent oxide thickness EOT = $t_{HfO2} (\varepsilon_{SiO2} / \varepsilon_{ox})$, where $\varepsilon_0$, $\varepsilon_{SiO2}$, $t_{HfO2}$, and $C_{HfO2}$ are the vacuum permittivity, silicon oxide dielectric constant, HfO$_2$ thickness, and HfO$_2$ capacitance respectively. We find dielectric constants of ~15, consistent with the high-k nature expected of ALD-grown HfO$_2$ films (Figure 3d).[46,49] These values translate to small EOTs of 2.8 and 4.4 nm for 11 nm and 16 nm thick HfO$_2$ films, respectively (Figure 3e). These measurements validate the potential of our technique to integrate high-quality, high-k HfO$_2$ with 2D materials while preserving excellent interface morphology with low trap density. High-k dielectric integration with atomically smooth surfaces free from interface traps is critical for high-performance 2D electronics. This enables efficient gate coupling to reduce SS towards the thermionic limit of 60 mV/dec while maintaining low gate leakage currents and minimal hysteresis for reliable low-power electronics.

To demonstrate this potential for electronic applications, we fabricated 2D WS$_2$ field effect transistors (FETs). The device architecture is schematically shown in Figure 4a and consists of an *n*-type WS$_2$ channel with 290 nm SiO$_2$/Si$^{++}$ as the global back gate and 11 nm HfO$_2$ as the top gate dielectric. An optical image of the device is shown in Figure 4b. Effective top-gate control is evident from the transfer curves where the drain current $I_D$ is measured while sweeping the applied top-gate voltage $V_{TG}$ (Figure 4c). The gate leakage current remains negligible, <1 pA/um over the entire gate range (shown in red). Our high-k HfO$_2$ facilitates efficient gate coupling, and we find a minimum SS value of 74.5 mV/dec, approaching the thermionic limit of 60 mV/dec (indicated by the blue slope for comparison). SS values remain low as the $I_D$ increases with negligible hysteresis between the forward and backward top-gate sweep (Figure 4d). Negligible hysteresis with a width of as small as ~3 mV at a back gate voltage $V_{BG}$ = 0 V is observed in the transfer curve sweeps due to the low interface trap densities expected for our devices (Figure 4e). More measurements showing negligible hysteresis for different sweep rates are shown in Supporting Information Figure S17. These findings are consistent with measurements of 5 other devices with similarly low SS values ranging from 74 - 115 mV/dec, low gate leakage currents from 2-6×10$^{-6}$ A/cm$^2$, and negligible hysteresis, as shown in Supporting Information Figure S19-24.

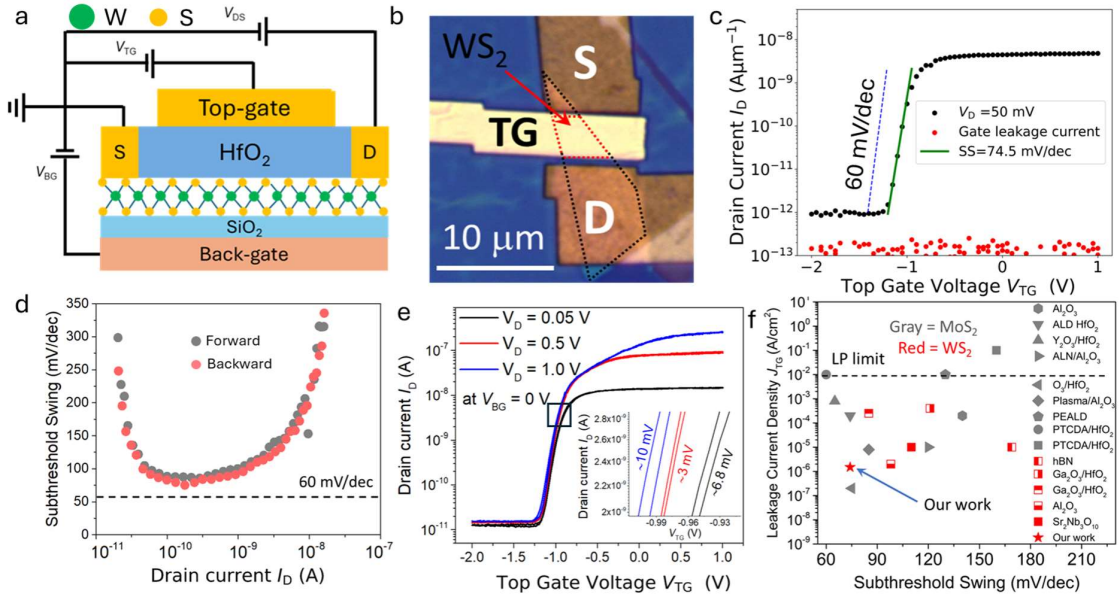

**Figure 4 Top-gated WS$_2$ transistor.** (a) Schematic of our transistor architecture with liquid metal-assisted transfer of HfO$_2$ gate dielectric. (b) Optical image of the device showing the source (S) and drain contacts (D) and the top gate electrode (TG) labeled. (c) Transfer curve of the device, the green line is the fit to extract the subthreshold slope *SS* of 74.5 mV/dec. (d) *SS* is determined as a function of the drain current for the forward and backward sweeps. (e) Drain voltage $V_D$ dependent transfer curves at back gate voltage $V_{BG}$ = 0 V showing negligible hysteresis at a sweep rate of 0.01 V/s. The inset shows a zoomed-in of the region indicated by the black square highlighting the hysteresis of transfer curves measured at different $V_D$. (f) Benchmarking of SS and gate leakage current density. Data for the gate leakage currents are measured at $V_D$=1 V. Due to the lack of data for WS$_2$ (red), we also include data for MoS$_2$ (gray), which typically shows better SS values due to lower interface trap densities from higher crystal quality. Our WS$_2$ transistor shows improved SS and gate leakage current values over other reported WS$_2$ works and is comparable with MoS$_2$ devices. More details of the benchmarking data including dielectric thickness is summarized in SI Table 1.

Finally, we critically assess the potential of our dielectric integration technique by benchmarking *SS* and gate leakage currents with other state-of-the-art approaches for single top-gated 2D transistors. Due to a lack of data available for WS$_2$ transistors, we also include MoS$_2$ devices in the benchmarking data. Note that while WS$_2$ is theoretically superior to MoS$_2$ for transistor applications due to its smaller effective mass, most experimental realization of 2D MoS$_2$ transistors outperforms WS$_2$ due to better material quality. As such, WS$_2$ transistors typically show higher *SS* values than MoS$_2$ transistors of similar architecture due to an increased interface trap density from intrinsic channel defects such as sulfur vacancies.[55] Nevertheless, our devices still exhibit competitive comparisons with MoS$_2$, and we demonstrate record-low *SS* for WS$_2$ transistors.

While we have focused primarily on LM gallium oxide, HfO$_2$, and WS$_2$, our approach is not material-exclusive. The transfer technique is generally compatible with other 2D vdW materials. Oxides that can be deposited with ALD or other physical vapor deposition techniques such as sputtering and evaporation should also be possible, e.g., Al$_2$O$_3$ and even HfZrO$_2$, which raises the intriguing potential for integrating ferroelectric thin films with 2D materials towards novel device architectures like negative capacitance transistors.[56] Likewise, materials beyond oxides can also be transferred with our approach, such as evaporated metals for contacts. A limitation in our work is that contact metals are first deposited before HfO$_2$ transfer. This results in a non-flat surface, leading to more significant gaps at the HfO$_2$ and 2D material interface. Possible solutions are using much thinner electrodes such as graphene or developing selective etching to enable HfO$_2$ to be transferred before contact fabrication. An even more powerful solution is to develop a single-step process to simultaneously integrate an entire prefabricated device stack that includes metal contacts and oxide dielectrics.[42] Such devices will have interfaces not compromised by multiple lithography steps during fabrication, which can degrade performance. This makes them attractive for the scalable fabrication of high-quality devices.

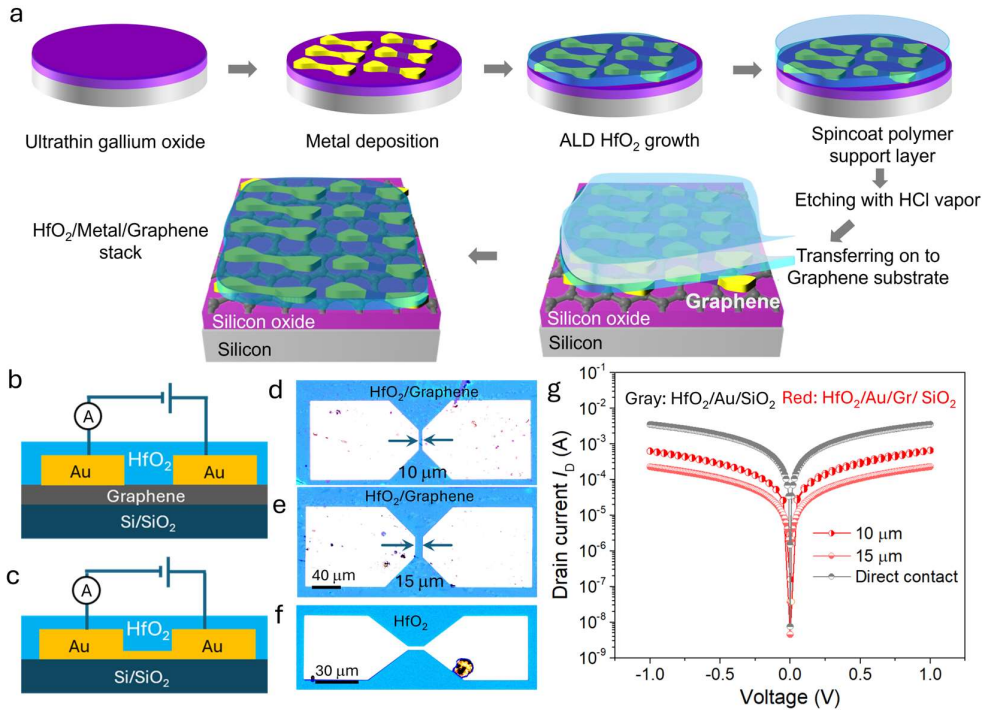

**Figure 5 Single-step transfer of dielectric and metal.** (a) Schematic illustration of our single-step transfer process of ALD HfO$_2$ and Au electrodes. Schematic of our devices made from transferring HfO$_2$ and (b) Au electrodes with varying gap separations onto graphene, or

(c) Au electrodes connected via an Au strip of length 25 µm and width 8 µm. (d-e) Optical images of the transferred Au electrodes with separation gaps of 10 and 15 µm transferred onto graphene. (f) Optical image of the transferred Au electrode connected via an Au strip. (g) Electrical measurements of the devices.

Here, we demonstrate such a proof-of-concept process by transferring simultaneously Au metal and HfO$_2$ dielectric in a single transfer step. Two types of geometries are shown. The first is Au pads with an area of ~120 µm$^2$ connected via an Au strip of length 25 µm and width 8 µm transferred to SiO$_2$ substrate (Figure 5a,c,f). We also transfer pairs of Au electrodes with varying separations of 10 and 15 µm (Figure 5b,d,e) onto chemical vapor deposition-grown graphene films. The mechanical integrity of the transferred Au is confirmed with electrical measurements of the connected Au electrodes showing a low resistance of ~300 Ω (Figure 5g). Measurements of the devices made from Au electrodes transferred on graphene show 2-10 kΩ resistances, indicating that the transferred Au electrodes can form good electrical contact with CVD graphene. These results demonstrate the potential for vdW integration of entire top-gated device stack arrays using our LM-assisted transfer approach.

Future work should target the optimization and scalability of our transfer process and explore its use for other materials. In principle, much thinner dielectric films should also be possible. Our main challenge working with thinner dielectric films was the poor optical contrast on SiO$_2$, which made visual identification difficult (Figure S9). This can be solved through appropriate substrate choices with refractive indices that can maximize optical contrast, as we showed with silicon nitride substrates (Figure 1b, S7c-d).[57] We also highlight that the LM synthesis approach has already demonstrated several different types of ultrathin oxides experimentally, and theory predicts that as many as fifty more may be possible.[52] This raises exciting new opportunities to experiment with other LM oxides as sacrificial transfer layers.

To conclude, we reported a LM-assisted integration of high-k ALD dielectric with 2D materials. LM oxides are remarkably suitable as a sacrificial layer to integrate high-quality ALD HfO$_2$ with 2D materials. WS$_2$ top-gated transistors fabricated with this approach show excellent performance with low gate leakage currents and record low subthreshold swings promising for low-power applications. We also demonstrate its potential for transferring entire prefabricated device stacks, paving the way toward the scalable integration of more complex circuits for applications in future nano-electronics.

**Methods**

**Atomic layer deposition:** $HfO_2$ thin films were deposited by atomic layer deposition (ALD) at 200 °C using argon as the purging and carrier gas, with tetrakis-ethylmethylaminohafnium (TEMAHf) and water precursors.

**Liquid metal gallium oxide printing:** A liquid gallium metal droplet of diameter ~1-3 mm is placed on a top transparent substrate. The top sapphire substrate (width = 1 cm, length = 2 cm) is attached to a micromanipulator to precisely position the liquid metal droplet over the target area where the ultrathin oxide film is intended to be printed. Next, the target substrate is heated up to ~60 °C and brought slowly into contact with the liquid metal droplet. The relative position between the substrates is slowly reduced (~1 µm/s) for the liquid metal droplet to thermally equalize with the heated bottom substrate. As the substrates are brought closer to each other, the liquid metal droplet is 'squeezed' and expands across the substrate surfaces, allowing the oxide skin to spread out and attach to the substrates. Finally, the substrates are brought apart at a rate of ~10 µm/s, leaving behind the ultrathin oxide skin with a roughness ~0.4 nm on the target substrate.[48] Next, excess liquid gallium residue can be mechanically cleaned by blue tape. Finally, the same sample is used to grow $HfO_2$ after a short $O_2$ plasma treatment.

**Liquid metal-assisted transfer of ALD $HfO_2$:** Transfer of smaller $HfO_2$ films is also possible using a polydimethylsiloxane (PDMS) stamp, which is convenient when fabricating vdW heterostructures based on micron-sized exfoliated flakes using a 2D material transfer station. The PDMS is first prepared by mixing the curing agent and base components in a 1:10 ratio. We then degas the mixture in a vacuum before pouring it onto a $SiO_2$ substrate to cure at 80 °C for 2.5 hours. This layer of PDMS is stripped and applied to a glass slide. $O_2$ plasma treatment at 100 W, 20 sccm $O_2$, and 200 mTorr for 5 s ensures effective adhesion between the glass and PDMS layer. We then performed a square cut on our PMMA/$HfO_2$/ $Ga_2O_3$ stack cut sample. The PDMS is attached to the PMMA/$HfO_2$/$Ga_2O_3$ stack and exposed to 1M HCl vapor in a closed vial at 60 °C for ~12h to selectively etch the gallium oxide. More details on the dependence of the etching process on molarity and duration can be found in the SI

Figure S2-3. Next, the etched stack is then immersed in DI water to release the $SiO_2$ substrate, followed by thorough water washing and vacuum drying to remove residual moisture. Finally, the $HfO_2$/PMMA/PDMS stack can be transferred onto target substrates using a 2D material transfer stage. Finally, the PMMA is removed with Microposit Remover 1165 solvent.

**Device fabrication:** Standard electron beam lithography defined contacts and pads. We utilized Nanofrazor thermal scanning probe lithography to define gate electrodes. 5/30 nm of In/Au was evaporated for contacts, and 5/35 nm of Cr/Au was evaporated for the gate electrodes.

**Au pads fabrication:** The Au pads depicted in Figure 5a were fabricated using electron beam lithography (EBL). Initially, a gallium oxide layer was printed on a $SiO_2$ substrate, followed by spin coating of a PMMA resist layer to define the desired pad patterns. Subsequently, Au was evaporated onto the substrate, and lift-off was performed using a 1165 solvent with an IPA rinse.

**Electrical measurements:** Electrical measurements were performed in an electrical probe station using Keithley 2450 SMUs.

**TEM and STEM imaging:** TEM specimens were prepared by cutting a thin slice of material across the exfoliated $WS_2$ crystal layer using a Thermo Fisher Helios 450s dual beam focused ion beam system. Imaging was performed with a Thermo Fisher Titan TEM, operated at 200 kV in bright-field TEM mode using an objective aperture that allows the first order diffracted beams of $WS_2$ to pass, defocusing the beam ~50 nm to further enhance the contrast.

**X-ray photoemission spectroscopy:** XPS data was acquired using an Al K$α$ source (photon energy $hv$=1486.7 eV) and a beam spot size of about 200 μm with an energy resolution of ≈0.3 eV. The photoelectrons were collected at a normal emission angle, and the light was incident at 60° to the surface normal. Binding energies were calibrated against the Au 4f core level energy of a gold reference sample. All XPS data was acquired at 300 K.

**Optical Measurements:** Room temperature photoluminescence (PL) and Raman spectroscopy were conducted using an Invia Raman Renishaw system. Point spectra were acquired with a 532 nm laser through a ×100 objective lens (numerical aperture,

NA = 0.85) and a 2400 lines/mm grating. To prevent any unintentional degradation from laser-induced heating, the excitation power was maintained below 50 µW. The sample mapping was performed by scanning a 15 × 10 µm area with the same laser excitation for both the pristine sample and the $MoS_2$ sample covered with transferred $HfO_2$.

## Acknowledgments

This research was supported by the Agency for Science, Technology, and Research (A*STAR) Grant C230917006, MTC YIRG grant No. M21K3c0124 and MTC IRG grant No. M23M6c0103. We acknowledge the funding support from Agency for Science, Technology and Research (#21709). K.E.J.G. acknowledges support from a Singapore National Research Foundation Grant (CRP21-2018-0094).